\documentstyle[12pt]{article}
\def\beq{\begin{equation}}
\def\eeq{\end{equation}}
\def\bea{\begin{eqnarray}}
\def\eea{\end{eqnarray}}
\def\la{\mathrel{\mathpalette\fun <}}

\def\fun#1#2{\lower3.6pt\vbox{\baselineskip0pt\lineskip.9pt
\ialign{$\mathsurround=0pt#1\hfil##\hfil$\crcr#2\crcr\sim\crcr}}}
\begin{document}

\def\be{\begin{equation}}
\def\ee{\end{equation}}
\def\sd{\strut\displaystyle}
\begin{flushright}
UAB-FT-352/94\\
\end{flushright}
\vskip 2 cm
\begin{center}
{\bf \Large
Standard Model predictions for\\ weak decays of $\eta$ mesons}
\vskip 2.5 cm
 A. Bramon \footnote{16414::BRAMON}\\
\vskip .1 cm
Grup de F\'\i sica Te\`orica, Universitat Aut\`onoma de Barcelona,\\
 08193 Bellaterra (Barcelona), Spain
\vskip .5 cm
and
\vskip .5 cm
E. Shabalin \\
\vskip .1 cm
ITEP, Moscow.
\vskip  4 cm
{\bf Abstract}
\end{center}
The branching ratios of weak decays  of $\eta$ mesons are estimated in the
framework of the Standard Model. To observe such decays, $\eta$
meson sources with $N_{\eta} > 10^{13}$ per year are needed.
\pagestyle{plane}
\setcounter{page}{1}
\newpage

The comparatively big number of $\eta$ mesons at the acting $\eta$ factory
(SATURN, $N_{\eta} \sim 10^{12}$ per year) and at future accelerators
(CELSIUS, $N_{\eta} \sim 3\cdot 10^9$ and DAPhNE, $N_{\eta} \sim
3\cdot 10^7$
per year) \footnote{For references see Table 5 in paper [1]} makes
pertinent the question on the
possibility to observe weak decays of $\eta$ mesons.

These decays possess some specific properties which would be good to
verify. Namely, because of $G$-parity conservation \cite{2}, \cite{3} the
decays \beq \eta\to \pi^{\pm}l^{\mp}\nu \eeq \beq \eta\to
\pi^0\pi^{\pm}l^{\mp}\nu \eeq are suppressed in comparison with the
analogous decays of $K^0_L$ mesons possessing the same CP-properties as
$\eta$.

The decays
\beq
\eta\to K^{\pm}e^{\mp}\nu
\eeq
are $G$-parity allowed, but they are suppressed compared to
$K^0_L\to\pi^{\pm}e^{\mp}\nu$ by the momentum-space factor $\sim
10^{-3}$.

Of the nonleptonic weak decays, the decay
\beq
\eta\to \pi\pi
\eeq
is the most interesting one
as it needs CP violation and it can not be masked by
strong interaction contributions as in the case of $\eta\to 3\pi$
decays.

Let's consider the above mentioned processes in detail.

\section{$\eta\to\pi^{\pm}l^{\mp}\nu$}

The hadronic part of the matrix element is
\beq
<\pi^{\pm}|J^1_{\mu}+iJ^2_{\mu}|\eta> =
f_+^{(\eta)}(q^2)(p_{\eta}+p_{\pi})_{\mu} +
f_-^{(\eta)}(q^2)(p_{\eta}-p_{\pi})_{\mu}
\eeq
In the Standard Model, where the second-class currents \cite{2} are absent,
the form factors $f_{\pm}$ can be different from zero only due to isospin
breakdown occurring through the virtual electromagnetic interactions or due
to the $m_d - m_u$ mass difference. As the last mechanism gives the
largest effect, one can estimate the probability of this decay
using a Chiral Theory of low-energy mesonic processes. The effective
Lagrangian approach, based on the idea that momentum dependence
of the form factors is determined mainly by spin 0 and spin 1
intermediate resonances \cite{4}, gives the result \footnote{
The form factor $f^{\eta}_{+}$ is originated by successive transitions
$\eta \to \pi^0 \to \pi^{\pm} l^{\pm}\nu$ with
$$
L(\eta \to \pi^0 ) = \frac{\sqrt{3}}{4} (m^2_{\eta} - m^2_{\pi})
\frac{m_d - m_u}{m_s - \frac{1}{2}(m_d+m_u)}
\cdot  (\cos \theta_P - \sqrt{2}\sin\theta_P)
$$
For $f^{(\eta)}_-$, the dependence on SU(3) breaking parameter is
really absent because the quantity $(m^2_\eta-m^2_\pi)$ itself is proportional
to
$m_s - \frac{1}{2}(m_d + m_u)$.}

\begin{eqnarray}
f_+^{(\eta)}(q^2)=-\sqrt{\frac{3}{8}}\frac{m_d -m_u}{m_s -\frac{1}{2}(m_d +
m_u)}[1+q^2/(M^2_{\rho} - q^2)] \cdot \\ \nonumber
\cdot  (\cos \theta_P - \sqrt{2}\sin\theta_P)
\end{eqnarray}
\begin{eqnarray}
f_-^{(\eta)}(q^2)=\sqrt{\frac{3}{8}}\frac{(m_d -m_u)(m^2_\eta-m^2_\pi)}
{m_s -\frac{1}{2}(m_d +
m_u)}[(M^2_\rho-q^2)^{-1}-(M^2_{a_0(980)}-q^2)^{-1}] \cdot \\ \nonumber
\cdot  (\cos \theta_P - \sqrt{2}\sin\theta_P)
\end{eqnarray}
where $\theta_P$ is the mixing angle in the pseudoscalar nonet.
Its prefered value \cite{4b} seems to be $\theta_P \simeq -19.5^{\circ}$
 from which one gets $\cos \theta_P - \sqrt{2}\sin\theta_P \simeq
\sqrt2$; other values (such as $\theta_P \simeq -10^{\circ}$) have also
been proposed, but our results are essentially independent of these
details.
\par
Therefore, the form factors of the decay (1) are suppressed by the factor
\beq
\beta=\sqrt{\frac{3}{8}}\frac{m_d -m_u}{m_s -\frac{1}{2}(m_d +
m_u)}\cdot ctg \theta_C  \cdot  (\cos \theta_P - \sqrt{2}\sin\theta_P)
\eeq
in comparison with the form factors of $K^0_L\to\pi^{\pm}l^{\mp}\nu$ decay.

Using the most conservative estimate \cite{5} for the quantity (8),
we come to the result
\beq
\beta\la 0.1
\eeq
Then
\begin{eqnarray}
B.r.(\eta\to\pi^{\pm}l^{\mp}\nu)\cong 2\beta^2(\frac{m_{\eta}}{m_K})^5
\frac{\Gamma(K^0_L\to\pi^{\pm}l^{\mp}\nu)}{\Gamma_{tot}(\eta)}
(\cos \theta_P - \sqrt{2}\sin\theta_P)^2 \\ \nonumber
\la 2\cdot
10^{-13} \cdot (\cos \theta_P - \sqrt{2}\sin\theta_P)^2
\end{eqnarray}

Therefore, an observation of the decay (1) with a rate considerably larger
than $10^{-13}$ would be an evidence on the existence of some new physics
beyond the SM. The estimates of possible contributions of the second-class
currents to this process are contained in refs. \cite{6} and \cite{7},
but the results are expressed in terms of unknown coupling constants of the
second-class current interaction.

\section{$\eta\to\pi^0\pi^{\pm}l^{\mp}\nu$}

The hadronic part of the matrix element is
\beq
<\pi^0\pi^+|A_{\mu}|\eta> = f_1(q^2)(p_{\pi}+p_{\pi'})_{\mu} +
f_2(q^2)(p_{\pi}-p_{\pi'})_{\mu} + f_3(q^2)
(p_{\eta}-p_{\pi}-p_{\pi'})_{\mu}
\eeq
\beq
<\pi^0\pi^+|V_{\mu}|\eta> =
f_4\frac{i\varepsilon_{\mu\nu\alpha\beta}}{M^2_K}(p_{\pi}+p_{\pi'})_{\alpha}
(p_{\pi}-p_{\pi'})_{\beta}(p_{\eta})_{\nu}
\eeq

Again, as in the case of the decay (1), the form factors $f_{1,2,3}$ are
suppressed by $G$-parity conservation and they are smaller than the
corresponding form factors for $K^0_L\to\pi^0\pi^{\pm}l^{\mp}\nu$ by
a factor $\beta$. This is not the case for the form factor $f_4$ which
is $G$-parity allowed. But the contribution of this form factor to the
$K^0_L\to\pi^0\pi^{\pm}e^{\mp}\nu$ decay rate is approximately 0.5\%.

Then the estimate for B.r. of $\eta_{e_4}$ decay is
\bea
B.r.(\eta\to\pi^0\pi^{\pm}l^{\mp}\nu) & \cong&  \\
\cong 2(\beta^2 \;\;
\mbox{\rm or} \;\; \frac{2}{3} ctg^2\theta_c \!\!\!\!\!\!
&\cdot&\!\!\!\!\!\! 0.005)
(\cos \theta_P - \sqrt{2}\sin\theta_P)^2 (\frac{m_{\eta}}{m_K})^7
\frac{\Gamma(K^0_L\to\pi^0\pi^{\pm}l^{\mp}\nu)}{\Gamma_{tot}(\eta)}\la
\nonumber \\
&\la& 1.7\cdot 10^{-16}\cdot (\cos \theta_P - \sqrt{2}\sin\theta_P)^2
\nonumber
\eea

\section{$\eta\to K^{\pm}e^{\mp}\nu$}

The hadronic part of the matrix element is of the form of (5). But
\beq
f_+(\eta\to K^- e^+\nu) = \sqrt{\frac{3}{2}}f_+(K^0\to\pi^-e^+\nu)
\eeq
and
\beq
f_-(\eta\to K^- e^+\nu) = \sqrt{\frac{3}{2}}(\frac{m^2_{\eta}-m^2_K}
{m^2_K-m^2_{\pi}})\cdot
f_-(K^0\to\pi^-e^+\nu)
\eeq
With these values of the form factors
\beq
\frac{\Gamma(\eta\to K^{\pm}e^{\mp}\nu)}{\Gamma(K^0_L\to\pi^{\pm}e^{\mp}\nu)}
\approx 0.867 \cdot 10^{-3}
\eeq
or
\beq
B.r. (\eta\to K^{\pm}e^{\mp}\nu)=4\cdot 10^{-15}
\eeq
with negligible $\eta-\eta'$ mixing effects.

\section{$\eta\to\pi\pi$}

Like $K^0_L\to 2\pi$ decay, this decay violates CP invariance \cite{8}, but
the strangeness does not change in $\eta\to 2\pi$ transitions.

For this reason, in the Standard Model, the amplitude of the
$\eta\to 2\pi$ decay must be suppressed at least by the factor
$G_F\Lambda^2 sin\theta_C$ (with $\Lambda\la 1$ GeV) in comparison with
the $K^0_L\to 2\pi$ amplitude. This estimate follows from the fact that CP
violation in SM occurs due to imaginary parts in the Yukawa couplings and to
have the observable effect of CP violation one needs to include
flavour-changing transitions like $\eta\to K^0\bar{K}^0$ containing
\underline{non-self-conjugated} product of the Yukawa couplings.
Considering then the decay of the $\{K^0, \bar{K}^0\}$ system
into $2\pi$ states
we come to the above estimate.

\section{Conclusions}

The estimates obtained in the framework of the Standard Model
$$
B.r. (\eta\to \pi^{\pm}l^{\mp}\nu)\la 2\cdot 10^{-13}
\cdot  (\cos \theta_P - \sqrt{2}\sin\theta_P)^2
$$
$$
B.r. (\eta\to\pi^0\pi^{\pm}l^{\mp}\nu)\approx 1.7\cdot 10^{-16}
\cdot  (\cos \theta_P - \sqrt{2}\sin\theta_P)^2
$$
$$
B.r. (\eta\to K^{\pm}l^{\mp}\nu)\approx 4\cdot 10^{-15}
$$
$$
B.r. (\eta\to 2\pi)\la 4[G_F\Lambda^2 sin\theta_C]^2
\cdot 10^{-14}
$$
show that
the observation of these decays at $\eta$  factories
with $N_{\eta} < 10^{13}$ per year would be the evidence of some new
physics beyond the Standard Model.

\section{Acknowledgments}

We are grateful to Prof. L.Okun for his indication that the first attempts
to estimate the weak widths of $\eta$ meson were undertaken in ref.
\cite{3}. He also attracted  our attention to the
recent paper \cite{9}, where
the same order estimate as ours was obtained for $\eta_{l_3}$ decay.

One of us (E.Sh.) takes great pleasure in acknowledging the hospitality
of the IFAE Theoretical Group at the Autonoma University of
Barcelona, where this work was fulfilled.

\end{document}